\documentclass[12pt]{article}

\usepackage{latexsym}
\usepackage{amssymb}
\usepackage{amsfonts}

\usepackage{hyperref}
\usepackage[dvips]{graphicx,color}
\graphicspath{{images/}}

\def\nc#1{\newcommand{#1}}
\def\rnc#1{\renewcommand{#1}}

\def\a{\alpha}
\def\b{\beta}
\nc{\g}{\gamma}
\def\d{\delta}
\nc{\D}{\Delta} 
\nc{\e}{\eta}
\nc{\ep}{\epsilon}

\nc{\ve}{\varepsilon}
\nc{\G}{\Gamma}

\nc{\la}{\lambda}
\nc{\La}{\Lambda}
\nc{\om}{\omega}
\nc{\Om}{\Omega}
\nc{\vphi}{\varphi}
\nc{\si}{\sigma}
\nc{\Si}{\Sigma}
\rnc\th{\theta}
\nc\Th{\Theta}
\nc{\z}{\zeta}

\def\psib{{\overline \psi}}
\def\phib{{\overline \phi}}

\def\thb{{\overline \theta}}

\def\cM{{\cal M}}
\def\cN{{\cal N}}
\def\cR{{\cal R}}

\nc{\got}[1]{\mathfrak{#1}} 

\def\det{{\rm det}}

\nc\im{{\rm Im}\, }
\nc\re{{\rm Re}\, }
\def\tr{{\rm tr}}

\def\exp{{\rm exp}}

\nc{\Rt}{{\tilde R}}
\nc{\CC}{{\mathbb C}}
\nc\II{{\mathbb I}} 
\nc{\RR}{{\mathbb R}}
\nc{\HH}{{\mathbb H}}
\nc{\NN}{{\mathbb N}}
\nc{\ZZ}{{\mathbb Z}}
\nc{\MM}{{\mathbb M}}

\def\dag{\dagger}

\nc{\eql}{\eqalign}
\nc{\dis}{\displaylines}
\nc{\ce}{\centerline}
\nc{\hf}{\hspace{\fill}}
\nc{\hs}{\hspace*}
\nc{\vs}{\vskip .3cm}
\nc{\non}{\nonumber\\}
\def\nn{\nonumber}

\nc{\noi}{\noindent}

\nc{\p}{\partial}
\nc{\na}{\nabla}
\def\x{\times}

\nc{\lb}{\left(}
\nc{\rb}{\right)}
\nc{\lan}{\langle}
\nc{\ran}{\rangle}

\nc{\beq}{\begin{equation}}
\nc{\eeq}{\end{equation}}
\nc{\beqa}{\begin{eqnarray}}
\nc{\eeqa}{\end{eqnarray}}
\nc{\beqas}{\begin{eqnarray*}}
\nc{\eeqas}{\end{eqnarray*}}
\nc{\barr}{\begin{array}}
\nc{\earr}{\end{array}}
\nc{\ben}{\begin{enumerate}}
\nc{\een}{\end{enumerate}}
\nc{\bit}{\begin{itemize}}
\nc{\eit}{\end{itemize}}
\nc{\ssec}[1]{\subsection{#1}}
\nc{\sssec}[1]{\subsubsection{\sc #1}}

\nc{\sfrac}[2]{{\mbox{\large $\frac{#1}{#2}\,$}}}
\nc\half{\sfrac{1}{2}}
\nc\nfrac[2]{\mbox{\small $\frac{#1}{#2}$}}

\nc{\eq}[1]{\stackrel{\ref{#1}}{=}}

\nc{\oto}{\leftrightarrow}

\def\refeq#1{{(\ref{#1})}}

\nc{\twovec}[2]{\left( \!\!
\begin{array}{c} #1\\  #2 \end{array}\!\!\right)}
\nc{\stwovec}[2]{\bigg( \!\!
\begin{array}{c} \mbox{\raisebox{.7ex}{$#1$}}\\[-.3cm]  #2
\end{array}\!\!\bigg)}
\nc{\twomat}[4]{\left(\!\! \begin{array}{cc} #1&#2\\ 
#3&#4\end{array}\!\! \right)}
\nc{\stwomat}[4]{\bigg(\!\!\barr{cc}
#1&\!\!\!\!#2\\[-.2cm]#3&\!\!\!\!#4
\earr\!\!\bigg)}

\makeatletter
\@addtoreset{equation}{section}

\makeatletter

\def\tE{{\widetilde E}}

\def\lab{\overline{\lambda}}

\def\lab{\overline{\lambda}}
\def\tla{\widetilde{\lambda}}
\def\mub{\overline{\mu}}
\def\mb{\overline{m}}
\nc\Wt{{\widetilde W}}

\begin{document}
\renewcommand{\thefootnote}{\alph{footnote}}
\setcounter{page}{0}

\begin{titlepage}
\titlepage
{\color{white}{x}}
\vspace{2cm}
\begin{center}
\LARGE{\bf Multi-Instanton Corrections to Superpotentials
  in Type II Compactifications\footnote{Work supported by  Polish Ministry of Science MNiSW 
under contract N202 176 31/3844 (2006-2008).}}
\end{center}

\vskip 1.5cm 

\begin{center}
{\large Tomasz Jeli\'{n}ski \footnote{\tt tjel@fuw.edu.pl} and Jacek Pawe{\l}czyk \footnote{\tt jacek.pawelczyk@fuw.edu.pl} }\\[0.5cm]
\textit{Institute of Theoretical Physics, University of Warsaw\\
Ho\.za 69, PL-00-681 Warsaw, Poland}
\end{center}

\bigskip

\begin{abstract}

We consider two simple examples of multi-instanton configurations in type II 4d $\mathcal{N}=1$ superstring compactifications. The first one involves $O(1)$ and $U(1)$ D2-instantons embedded in $T^6/(\ZZ_2\x\ZZ_2')$ geometry with $SU(4)$ gauge symmetry coming from D6-branes and the second is related to quiver gauge theory of orientifolded orbifold of conifold containing fractional D(-1)-instantons and D3-branes.  The additional zero modes of instantons which appear at the intersection points are lifted through interactions stemming from dimensionally reduced F- and D-terms. It is shown that multi-instanton configurations effectively generate corrections to the usual non-perturbative superpotential for chiral matter.  

\end{abstract}

\vfill
\begin{flushleft}
{\today}\\
\end{flushleft}
\end{titlepage}

\newpage
\tableofcontents

\newpage
\setcounter{footnote}0
\renewcommand{\thefootnote}{\arabic{footnote}}
\section{Introduction}
Instantons are necessary ingredients of string theory providing non-perturbative corrections to string interactions
among which the most important are corrections to the superpotential.
These were used e.g. to stabilize K\"ahler moduli \cite{kklt}, break SUSY \cite{aks} and influence the inflaton motion \cite{d3-inst-conif}.
D-brane instanton effects has been extensively studied in various explicit compactification models either in type IIA \cite{seesaw,majorana,lifting,yukawa} or IIB string models \cite{argurio,bianchi}. 
Until very recently it has been widely believed that among various possible D-instantons the 
only relevant are the one-instanton configurations
 \cite{Witten96,Beasley,seesaw,majorana,lifting,yukawa}. 
As it turns out some multi-instantons\footnote{\cite{argurio,bianchi,petersson,nonBPS}} can in fact generate superpotential \cite{petersson,uranga}. 

In this paper we shall follow this procedure and investigate 
several multi-instanton configurations renormalizing superpotentials. 
The crucial role is played by zero modes of instantons and their mutual interactions. Thus the first step in analysis of any D-brane instanton configuration is to establish what kind of zero modes are present. Then one has to find effective action for these modes, integrate over all of them and check whether due to interactions path integral is non-zero. If it is nontrivial then one can seek effects in 4d theory generated by D-brane instantons. 
In the type IIA context we shall discuss
 two D2-instantons in orientifolded geometry $T^6/(\ZZ_2\times\ZZ'_2)$ \cite{frozen} with D6-branes. For the type IIB superstring we shall estimate the contributions of several
 fractional D(-1)-instantons placed in nodes of the quiver diagram of the orientifolded $\ZZ_n$ orbifold of conifold \cite{aks} containing fractional D3-branes.

The paper is organized as follows. In section 2 we recall what kind of zero modes appear in discussed systems and what are interactions occurring in multi-instanton configurations. We describe them by means of effective action. Section 3 includes two simple examples (one in type IIA and one in type IIB superstring theory) of multi-instanton configurations which we analyse using effective action obtained in section 2.  Appendix A contains computations and details of geometry underlying type IIA example of multi-instanton configuration.

\newpage
\section{Instantons zero modes and their interaction}
\label{e-inst}

Here we consider orientifolded type II superstrings  compactified on $\RR^{3,1}\times Y$, where $Y$ is Calabi-Yau.
Generic BPS D-branes (or Euclidean D-instantons)  preserve half of bulk supercharges breaking $\cN=2$ to $\cN=1$ thus  having four fermionic zero modes $\th$, $\thb$.  When they do not have extra fermionic zero modes they are called rigid\footnote{Non-rigid instanton have extra fermionic zero modes coming e.g. from gauge degrees of freedom: these enter the so-called `McLean multiplets'.
Such instantons will not contribute to the superpotential.} $U(1)$ instantons due to $U(1)$ gauge symmetry of its world-volume \cite{lifting}. These modes are accompanied by bosonic fields thus together they  form the following multiplet\footnote{It is dimensionally reduced vector multiplet of the D6-brane.} 
\beq\label{vec-m}
(X^\mu,\ \th,\ \thb,\ D).
\eeq

It is known that corrections to superpotential are generated by instantons with only two fermions zero modes $\th$. In the present setup  they are invariant under orientifold projection and are called $O(1)$ instantons \cite{lifting}. The orientifolding kills half of the fermionic states of \refeq{vec-m} yielding the multiplet\footnote{The projection for anti-instanton will kill $\th$.}
\beq\label{vec-m-o}
(X^\mu,\ \th,\ D).
\eeq

In this paper we focus on multi-instanton configurations intersecting matter D-branes and intersecting each other. 
D-branes intersections provide matter while intersections with instantons produce extra zero modes.   The modes are bifundamentals of the gauge group of intersecting branes/instantons. For the instanton-instanton intersection away from any O-plane (regardless if it is $U(1)$ or $O(1)$ instanton) the modes are 
\beq\label{chiral}
\Xi=(\phi,\ \psi,\ F)
\eeq
and their c.c. fields \cite{nonBPS}. The multiplet \refeq{chiral} is dimensionally reduced chiral multiplet appearing at the analogous D6-D6 brane intersection \cite{berkooz}. 

We also have to include the possibility that $U(1)$ instanton intersects its orientifold image on top of an orientifold plane.
Resulting extra zero modes are:
$\half(\Pi_{E'}\circ \Pi_{E}+\Pi_O \circ \Pi_{E})$ modes $(m,\mb,\mub)$ and
$\half(\Pi_{E'}\circ \Pi_{E}-\Pi_O \circ \Pi_{E})$ modes $\mu$, where $(m,\mb)$
are bosonic  while $(\mu,\mub)$ fermions variables
\cite{lifting,nonBPS}. 

The spectrum changes also for the brane-instanton intersection. The technical reason is that in Minkowski directions open strings have mixed ND boundary conditions. A careful treatment of the BCFT shows (\cite{seesaw}) that there is only one fermionic zero mode $\la$ coming from the Ramond sector and  its charge depends on the intersection number $\Pi_E\circ\Pi_a$ ($\Pi_{E,a}$ denote cycles occupied by instanton and branes respectively):
\beq\label{lambda}
\la\in (\square,-1)\quad \textrm{when}\quad\Pi_E\circ\Pi_a>0.
\eeq 
When instantons intersect also orientifold image of D6-branes, then the image zero mode\footnote{Modes $\lab'$ (i.e. image of modes $\la$) correspond to intersection of instanton and image of branes. Thus if $\Pi_E\circ\Pi_a>0$, then in the case of $O(1)$ instanton one gets $\Pi_E\circ\Pi'_a=-\Pi_E\circ\Pi_a<0$. This implies that $\lab'\in(\square,1)$ \cite{lifting}.} $\lab'$ transforms as  $(\square,1)$ \cite{lifting}. Because on  $O(1)$ instanton charges $+1$ and $-1$ are equivalent we can identify $\lab'=\la$.

One must remember that globally, the charges ($U(1)_E$ or $\ZZ_2$) of the instanton
zero modes  always sum up to zero.

\ssec{Interactions}

What makes the instanton contributions non-trivial is the interaction of the instanton zero modes with matter and between themselves. The latter leads to the reduction of their number. 

For all the cases relevant in this paper all the interactions originate from intersections and they are of two types.
Local D-terms\footnote{The zero modes $\la$ are not $SO(4)$ spinors thus they do not enter D-terms.} come  from the standard gauge invariant interaction for $\cN=1$ chiral multiplet \refeq{chiral} with vector multiplet \refeq{vec-m} or \refeq{vec-m-o} \cite{WessBagger}
\beqa\label{d-term}
\half D^2+|F|^2-i\frac{q}{\sqrt2}(\phi\,\psib\,\thb-\phib\psi\,\th)
+\half (q\,D+\half q^2 (X^\mu)^2)\phib\phi+\xi D.
\eeqa
In the following we shall put the FI term $\xi=0$. Thus e.g. at the $U(1)$ instanton-instanton intersection
\beqa\label{ee}
S_{EE}&=&\half( D^2_1+D_2^2)+|F|^2-i\frac{q}{\sqrt2}(\phi\,\psib\,(\thb_1-\thb_2)-\phib\psi\,(\th_1-\th_2))\nn\\
&&+\half (q(D_1-D_2)+\half q^2(X_1^\mu-X_2^\mu)^2) \phib\phi.
\eeqa
The last term forces us to set $X_1=X_2$ to ensure that instantons intersect each other, otherwise strings connecting them are heavy. If one of the two instantons is $O(1)$ (say the one  subscripted with 1),  then one should truncate $\thb_1-\thb_2$ to $-\thb_2$. 
One can analogously construct the action $S_{EE'}$ for zero modes  which are localized at the intersection of $U(1)$ instanton and its orientifold image. 

\begin{figure}[!h]
\begin{center}
\includegraphics{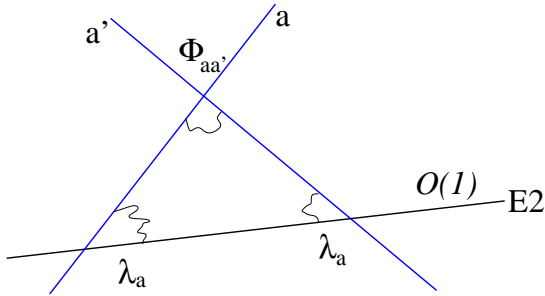}
\caption{Non-local interaction between E2-brane (black line) and two D6-branes (blue lines).}
\label{dde}
\end{center}
\end{figure}

The other interactions come from F-terms. Consider two intersecting D6-branes\footnote{D$6'_a$-brane is orientifold image of D$6_a$-brane.}: D$6_a$ and D$6'_a$ intersected by an $O(1)$ D2-instanton (i.e. E2-brane) as depicted in figure \ref{dde}.
\noindent The string disc diagram gives the following coupling\footnote{Here we identify $\la$ and $\lab'$.} \cite{seesaw}
\beq\label{llphi}
g_{e} \la_a \Phi_{aa'}\la_a,
\eeq
where $\Phi_{aa'}$ is the chiral supermultiplet and $g_e=g_s e^{-A}$ for $A$ being the area of the minimal world-sheet spanned between intersection point and having boundary on branes. Notice that the net charge (under the E2 gauge group) of the 
 fermionic zero modes appearing in \refeq{llphi} must vanish. Also,
\refeq{llphi} has formally the same structure as the superpotential obtained if one would replace the E2-instanton by the D6-brane (subscripted here by $c$) i.e.
\beq\label{abc}
W_{aa'c}= g_e \tr(\Phi_{ca}\Phi_{aa'}\Phi_{a'c}).
\eeq
This can be easily generalized to other known superpotentials for D-brane matter e.g. those obtained in quiver gauge theories (see section \ref{conifold}).
Integrating out $\la$'s results in
\beq\label{sup0}
\d W=\mathrm{Pf} (g_e \Phi_{aa'})\ e^{-\widetilde V}.
\eeq
where  $\widetilde V$ is the volume of the instanton.
It is obvious that  $\mathrm{Pf}(g_e \Phi_{aa'})$ naively breaks $U(1)_a$ symmetry. The mechanism of its restoration goes through generalized Green-Schwarz mechanism and it involves an axion contained in $\widetilde V$.\\

\section{Multi-instanton configurations}

In this section we are going to describe several multi-instanton configurations contributing to superpotentials.
We hope that discussed examples will be illustrative enough to convince the reader that multi-instantons generically do contribute to superpotentials and that the contributions might be  interesting for applications in more realistic string-inspired models of unification.  

First we concentrate on orientifolded orbifold $T^6/(\ZZ_2\times\ZZ_2')$ with single stack of D6-branes. The model has $U(4)$  gauge symmetry and 32 chiral multiplets in anti-symmetric representation\footnote{Diagonal $U(1)\subset U(4)$ is anomalous and the related gauge boson gets mass via Green-Schwarz mechanism. Moreover, matter content of this model implies that $SU(4)$ is IR-free.}. The second example is a quiver theory stemming from orientifolded $\ZZ_n$ orbifold of conifold. Both models are simple enough to  exhibit all necessary ingredients of the multi-instanton calculus.
 
Few remarks are necessary before we set to concrete calculations. First of all, any nontrivial instanton configuration must contain one $O(1)$ instanton. Also,  due to the $\mathcal{R}$ non-invariance\footnote{$\mathcal{R}: z^i\mapsto\bar {z}^i$ on $i$-th two-torus $(T^2)^i$; it is part of the orientifold action $(-1)^F\Om\mathcal{R}$.} any $U(1)$  instanton occurs with its image. We expect that the number of $U(1)$ instantons contributing  to the superpotential can be arbitrary large. 
Each of the instantons has its own fermionic zero modes. Reduction of their number is due to instanton-instanton interaction and sometimes it is called recombination \cite{lifting}.

There are two basic types of multi-instanton configurations of interest: either the extra $U(1)$ instantons do or do not intersect D-branes. In the latter case the resulting contribution renormalizes the result by extra dependence on closed string moduli. From the point of view of phenomenology  this seems to be not very interesting. On the other hand when the $U(1)$ instantons intersect D-branes they may lead to non-trivial corrections to $W$.  As we shall see they are of the form $\mathrm{Pf}(\Phi_{aa'})\det(\Phi_{aa'})^k$ and may significantly influence physics of some models.

\subsection{Multi-instantons in $Y=T^6/(\ZZ_2\times\ZZ_2')$ model}
Here we shall consider orientifolded compactification model $Y=T^6/(\ZZ_2\times\ZZ_2')$ containing one stack of four D6-branes\cite{frozen,seesaw,majorana}. Details of the underlying geometry and some calculations can be found in appendix A. 

\subsubsection{U(1) instantons not intersecting D6-branes}\label{twonotint}
\begin{figure}[h]
\begin{center}
\includegraphics{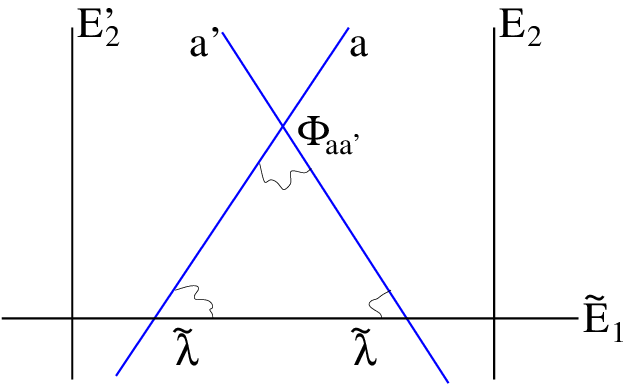}
\caption{}\label{IIA-3inst}
\end{center}
\end{figure}

Consider brane-instanton configuration presented in figure \ref{IIA-3inst}. Beside $O(1)$ D2-instanton (denoted by $\widetilde{E}_1$) it contains $U(1)$ D2-instanton (denoted by $E_2$) and its orientifold image (denoted by $E'_2$) both not intersecting D6-branes, 
$\widetilde{E}_1$ intersects D6-branes once thus here occurs one multiplet $\mathbf{4}$ of $SU(4)$  fermionic modes $\widetilde{\la}_i$, $E_2$ intersects $\widetilde{E}_1$ giving modes \refeq{chiral}. Their number  depends on $E_2$
wrapping numbers. The leading contributions are those with the least wrapping number. In the  case at hand  it occurs when $E_2$ intersects $\widetilde{E}_1$ only once (see appendix \ref{app-IIA}).      
Then one can integrate over the localized degrees of freedom $(\widetilde{\la}_i,\Xi, \Xi^\dag)$ and over \refeq{vec-m}.  Under  the integral one can  shift $\th_1-\th_2\to -\th_2$. In this way all fermions (beside $\th_1$) can be soaked up yielding the following superpotential(details of computations can be found in appendix A)\footnote{If $M=(M_{ij})$ is $2n\times 2n$ skew-symmetric matrix then $\mathrm{Pf}(M)=\sum_{\si\in S_{2n}}\mathrm{sgn}\,\si\prod_{i=1}^nM_{\si(2i-1),\si(2i)}$}:
\beq\label{sup00}
 \a\, \mathrm{Pf}(g_e^I\Phi_{aa'}^I)e^{-\widetilde{V}-V},
\eeq
where $\a$ is some non-zero constant, $V$ is volume of $U(1)$ D2-instanton and $\widetilde{V}$ is volume of $O(1)$ D2-instanton.

\subsubsection{Instantons intersecting D6-branes} \label{twoint}
In a sense the result \refeq{sup00} is not very interesting because it only renormalizes the known one-instanton
 contribution \refeq{sup0}. The question is if multi-instanton configurations can produce  contributions of different type. In this section we shall show the answer is positive producing terms of the form $\mathrm{Pf}(\Phi_{aa'})\det(\Phi_{aa'})^k$. In order to get contributions of this type we need $U(1)$ instantons
intersecting D6-branes.

The analysis of the full set of three-instanton contributions is too complicated to be presented here thus we limit the considerations to  some of the non-vanishing terms.  Thus we display only those  configurations which  will contain figure \ref{dde} as a subdiagram.
As a side remark we notice that this is not the only possibility. In some cases the non-vanishing disc diagram can have more than three vertex operators  thus of higher order in string coupling constant. Its explicit value might be difficult to get. We shall not discuss such contributions here.

Because we focus on this specific set of possibilities we require 
that $E_2$ intersects: 
\ben
\item[a)] D6-branes with positive orientation $\Pi_2\circ\Pi_a>0$ - this gives $L_{\la}:=\Pi_2\circ\Pi_a $ modes $\la$,
\item[b)] D$6'$-branes with negative orientation $\Pi_2\circ\Pi'_a<0$ - here one gets $L_{\lab'}:=-\Pi_2\circ\Pi'_a$ modes $\lab'$,
\item[c)] $\tE_1$ such that $\Pi_2\circ\widetilde{\Pi}_1>0$ - here one gets $L_\Xi:=\Pi_2\circ\widetilde{\Pi}_1$ multiplets $\Xi$ \refeq{chiral} and $\Xi^\dag$; these have -1 and +1 $U(1)_{E_2}$ charge respectively, 
\item[d)] $E'_2$ and O6-plane such that $\half(\Pi_2'\circ\Pi_2+\Pi_O\circ\Pi_2)>0$ and $\half(\Pi_2'\circ\Pi_2-\Pi_O\circ\Pi_2)>0$; we assume that $E_2$ intersects $E'_2$ and O6-plane so we have to take into account zero modes $m$, $\mu$, $\overline{m}$ and $\mub$ localized at those intersections \cite{nonBPS}. Note that
due to cancellation of total $U(1)_{E_2}$ charge\footnote{The total $U(1)_{E_2}$ charge of $m$, $\mu$, $\overline{m}$ and $\mub$ must cancel the total charge of $\la$'s and $\lab'$'s.} and orientifold projection\footnote{Appendix A in \cite{lifting}.} one gets $L_{\mub}:=\half(\Pi_2'\circ\Pi_2+\Pi_O\circ\Pi_2)$ modes $m$, $\mb$ and $\mub$ and $L_\mu:=\half(\Pi_2'\circ\Pi_2-\Pi_O\circ\Pi_2)$ modes $\mu$. $m$ and $\mu$ have $+2$ $U(1)_{E_2}$ charge; $\overline{m}$ and $\mub$ have $-2$ charge.
\een
These determine which type of interactions occur. The easiest way to encode them is through  a quiver diagram.
In the quiver diagram we shall distinguish representations $(\mathsf{rep}_{U(4)},\mathsf{rep}_{U(1)})$ as follows. We assign arrowheads to lines connecting gauge groups: if the $U(1)$ charge of field is +1 then one draws $ -\!-\!\!\!<\bullet$, where $\bullet$ corresponds to $U(1)$ group; if the charge is -1 then one draws $-\!-\!\!\!>\bullet$. For given node an interaction is gauge invariant iff $-\!-\!\!\!<$ and $-\!-\!\!\!>$  match to each other i.e.  at this node one has:
\beq 
-\!\!\!-\!\!\!<\bullet<\!\!\!-\!\!\! -\quad\textrm{or}\quad -\!\!\!-\!\!\!>\bullet>\!\!\!-\!\!\! -.
\eeq
There is one more rule which concerns $O(1)$ groups -- for these we draw rectangular node instead of circular one and there is no arrowhead at the end of line touching rectangular nodes.

%
%
\begin{figure}[!h]
\begin{center}
\includegraphics[scale=0.4]{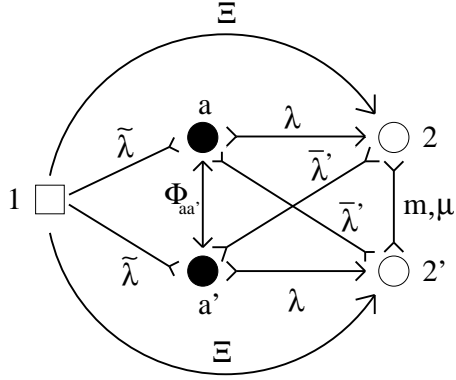}
\end{center}
\caption{Quiver diagram for intersecting D6-branes and D2-instantons: rectangle denotes $O(1)$ instanton, empty circles denote $U(1)$ instanton and its orientifold image and solid circles denote D6-branes and their images. Hermitean conjugates of $\Xi$, $m$ and $\mu$ are not displayed on the diagram.}\label{quiver}
\end{figure}

Using rules enumerated above it is easy to draw quiver diagram which corresponds to system satisfying conditions a) - d)  (see figure \ref{quiver}) and to list all (trilinear and potential) interactions occurring in this system\footnote{The couplings to $\th_2$ and $\thb_2$ come  from D-terms for $U(1)_{E_2}$ gauge theory -- see section 2.}:
\beq
\tla\Phi_{aa'}\tla+\la\Phi_{aa'}\lab'+m\psi\,\psi+\mu\psi\phi+\mb\psib\,\psib+\mub\psib\,\phib+m\mub\thb_2+\mb\mu\th_2+\phi\psib\,\thb_2
+\phib\psi\th_2+U,
\eeq
where we have suppressed gauge indices, denoted sum over all modes $\la_J$ by $\la$, sum over all modes $m_I$ by $m$  etc.\footnote{Indices $I$, $J$, etc. count intersection points.}, and incorporated coupling constants into field definitions. $\phi_{aa'}$ is bosonic component of chiral field $\Phi_{aa'}$. $U$ is potential coming from dimensionally reduced F- and D-terms.

So now the task is to integrate over all fermionic zero modes. One can see that integration over $\tla$ gives factor $\mathrm{Pf}(g_e^I\Phi_{aa'}^I)$. Moreover the necessary condition for non-vanishing of path integral $Z$ is $L_\la=L_{\lab'}$. If this is fulfilled then  
\beq
Z=\mathrm{Pf}(g_e^I\Phi_{aa'}^I)[\det(g_e^I\Phi_{aa'}^I)]^{L_\la}\int d\mathcal{M}\,e^{-S},
\eeq
where $d\mathcal{M}$ is the measure containing zero modes other than $\tla$, $\la$ and $\lab'$.
It turns out that integration over fermionic zero modes gives the following expression:
\beq\label{neutral}
\int d\phi\, d\phib\,dm\,d\mb\, \phi^\a\phib^\b m^\g\mb^\d F(|\phi|,|m|),
\eeq
where $\a$, $\b$, $\g$, $\d$ are integers and $F$ is a function which depends only on $|\phi|$ and $|m|$. So $Z$ is non-vanishing iff $\a=\b$ and $\d=\g$.  Then the integral over bosonic modes is just a non-zero constant $C$ and the contribution to the superpotential is $\d W=C\, \mathrm{Pf}(g_e^I\Phi_{aa'}^I)[\det(g_e^I\Phi_{aa'}^I)]^{L_\la}e^{-\widetilde{V}-V}$. One can show that for described geometry  $\a=2L_\mu -2$, $\b=2L_{\mub} -2$, $\g=L_\Xi -L_\mu -1$ and $\d=L_{\Xi} -L_{\mub} -1$. 

The next step is to express conditions $\a=\b$, $\g=\d$\footnote{It turns out that condition $L_\la=L_{\lab'}$ is equivalent to $L_\mu=L_{\mub}$.} and a) - d) in terms of wrapping numbers $(n^i,m^i)$:
\beqa 
&&m^1(n^2n^3+m^2m^3)=n^1(n^2m^3+m^2n^3)\nn\\
&&n^1(n^2n^3+m^2m^3)<m^1(n^2m^3+m^2n^3),\quad n^1m^2m^3>0,\quad m^1n^2n^3>0.
\eeqa
These (in)equalities describe big class of D2-instantons but contributions to superpotential $W$ are not equal for all of them. The  leading contribution to $W$ comes from instantons having minimal volume $V$:
\beq
V\sim\prod_{i=1}^3[(n^i)^2+(m^i)^2]^\frac{1}{2}.
\eeq
These  are:
\beqa
&&n^1=4,m^1=5,n^2=1,m^2=2,n^3=2,m^3=1\nn\\
&&n^1=-4,m^1=-5,n^2=1,m^2=2,n^3=-2,m^3=-1\nn\\
&&n^1=-4,m^1=-5,n^2=-1,m^2=-2,n^3=2,m^3=1\nn\\
&&n^1=-4,m^1=-5,n^2=-1,m^2=-2,n^3=-2,m^3=-1
\eeqa
For such configurations the contribution to the superpotential is
\beq\label{WUO}
\d W(\Phi_{aa'},\widetilde{V},V)=C\,\mathrm{Pf}(g_e^I\Phi_{aa'}^I)[\det(g_e^I\Phi_{aa'}^I)]^9e^{-\widetilde{V}-V}.
\eeq

\subsection{Orientifolded $\ZZ_n$ orbifold of conifold and D(-1)-instantons}
\label{conifold}

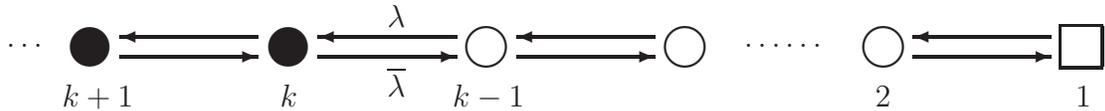
\begin{figure}[ht]
\begin{center}
\setlength{\unitlength}{0.75pt}
\begin{picture}(150,65)(-240,-25)
\thicklines\put(-400,0){\circle*{20}}
\thicklines\put(-300,0){\circle*{20}}
\thicklines\put(-200,0){\circle{20}}
\thicklines\put(-210,-10)
\thicklines\put(-100,0){\circle{20}}
\thicklines\put(-110,-10)
\thicklines\put(0,0){\circle{20}}
\thicklines\put(-10,-10)
\thicklines\put(90,-10){\framebox(20,20)}
\thicklines\put(-385,-5){\vector(1,0){70}} 
\thicklines\put(-315,5){\vector(-1,0){70}} 
\thicklines\put(-285,-5){\vector(1,0){70}} 
\thicklines\put(-215,5){\vector(-1,0){70}}
\thicklines\put(-185,-5){\vector(1,0){70}}
\thicklines\put(-115,5){\vector(-1,0){70}}
\thicklines\put(15,-5){\vector(1,0){70}} 
\thicklines\put(85,5){\vector(-1,0){70}}
%
%
%
\put(-442,0){\ldots}
%
%
\put(-70,0){\ldots\ldots}
\put(97,-30){1}
\put(-4,-30){2}
\put(-217,-30){$k-1$}
\put(-304,-30){$k$}
\put(-414,-30){$k+1$}
\put(-250,10){$\la$}
\put(-250,-25){$\lab$}
\end{picture}
\caption{Quiver diagram for orientifolded $\ZZ_n$ orbifold of conifold. Filled circles denote D3-branes, empty circles denote $U(1)$ instantons and square denote $O(1)$ instanton.}
\label{conf-z}
\end{center}
\end{figure}

Now we shall turn to IIB superstring on conifold singularity with some D3's. The model of interest is its orientifolded $\ZZ_n$ orbifold. For details we refer to \cite{aks}.
 The quiver diagram of such a theory is depicted in Fig. \ref{conf-z}. 
The quiver diagram encodes the matter content of the theory but also the superpotential $W=\sum_{i=2}^{n} W_i$, where
\beq\label{Wfourphi}
W_i=(-1)^i h\ \Phi_{i-1,i}\Phi_{i,i+1}\Phi_{i+1,i}\Phi_{i,i-1}.
\eeq
Here $\Phi_{i,j}$ denote chiral superfields which occur when nodes are occupied by D3-branes. When two adjacent nodes are occupied by D(-1)-instantons then one gets the multiplet $\Xi$ \refeq{chiral} instead of $\Phi_{i,j}$.  
The gauge group at each node is $U(k_i)$ except the first one for which the group is $USp(k_1)$ (for $O3^-$-plane and $SO(k_1)$ for $O3^+$-plane). From now on we assume that for all nodes $i=1,\ldots,k$: $k_i=1$.
  Let all nodes starting from the first one till the $(k-1)$-th be occupied by one instanton . The system has also zero modes $\Xi$ 
corresponding to bifundamentals of the instanton gauge groups.
These zero modes interact among themselves with superpotential as in \refeq{Wfourphi} with appropriate field substitutions (i.e. $\Phi_{i,j}\longrightarrow\Xi_{i,j}$). 
The only important difference is that the instanton at the $(k-1)$-th node has the exotic $W$ \cite{aks,simic}
\beq\label{Wla}
h (\la\,\Phi_{k,k+1}\Phi_{k+1,k}\lab-
 \la\,\Xi_{k-1,k-2}\Xi_{k-2,k-1}\lab)
\eeq
due to exotic zero modes arising from D-brane-instanton intersection 
as described in Sec.\ref{e-inst}. Notice that the instanton does not provide anti-holomorphic analog of  \refeq{Wla}
(these come from anti-instantons). 

The simplest $k=2$ case was considered in \cite{aks}.
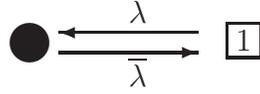
\begin{figure}[!ht]                                                
\begin{center}                                                      
\setlength{\unitlength}{0.75pt}                           
\begin{picture}(150,80)(-130,-25)
\thicklines\put(-100,0){\circle*{20}}
\thicklines\put(0,-7){\framebox(16,16){1}}
\thicklines\put(-85,-5){\vector(1,0){70}}
\thicklines\put(-15,5){\vector(-1,0){70}}
\put(-50,10){$\la$}
\put(-50,-22){$\lab$}
\end{picture}\label{quiverO}
\caption{Quiver diagram for $k=2$.}
\end{center}
\end{figure}
From \refeq{Wla} one gets the coupling $\la\lab\Phi_{23}\Phi_{32}$. Integrating over $\la$'s gives $\d W~\propto~ h\Phi_{23}\Phi_{32}$.

For $k>2$ we have plenty of zero modes coming from extra instantons, but most of them are not really zero modes due to interactions discussed in the previous section. The integration over $\la$, $\lab$ goes as before. In the path integral over the other zero modes there is no $\thb_1$ due to $USp(1)=O(1)$ at the first node. The dependence on the other $\th$'s is through $\th_i-\th_{i+1}$,  $\thb_i-\thb_{i+1}$ only, thus one can shift
$\th_i-\th_{i+1}\to \th_{i+1}, \ \thb_i-\thb_{i+1}\to \thb_{i+1}$. This allows to integrate over $\th$'s producing
\beq\label{zero-t}
\int \prod_{k=1}^{k-2}d\phi_{i,i+1}d\phib_{i,i+1}d\phi_{i+1,i}d\phib_{i+1,i}|Z|^2 e^{-S_\phi},
\eeq
where  $S_\phi$ is an action for bosonic zero modes only
\beq
S_\phi = \sum_{i=1}^{k-2}\left(|\phi_{i,i+1}|^2-|\phi_{i+1,i}|^2\right)^2+S_F
%
\eeq
where $S_F\propto\phi^6$  denotes the F-term contribution to the potential for $\phi$'s  coming from $W$
and $Z$ is given by 
\beq\label{h-ah}
\sim \left|\prod_{i=1}^{k-2}d^2\psi_{i,\,i+1}d^2 \psi_{i+1,\,i}(\phib_{i,\,i+1}\psi_{i,\,i+1}-\phib_{i+1,\,i}\psi_{i+1,\,i})^2\ \exp(-\Wt|_{\th\th})\right|^2, 
\eeq
where $\Wt$ is the superpotential for zero modes without terms with $\la$'s i.e. \refeq{Wla} and without potential terms for bosonic modes (these terms are included in $S_\phi$).
Thus the integral \refeq{zero-t} vanishes iff $Z=0$. Notice that   in \refeq{h-ah} one integrates over holomorphic fermionic zero modes only.

We need to analyse $Z$. It is clear that for $Z\neq 0$ one needs contributions from $\Wt$. Thus $Z=0$ for $k=3$. As we shall see this is exceptional case. For other $k$'s $Z\neq 0$.  We simplify the task choosing a special point at the space of the bosonic zero modes. Let us first consider even $k$. 
Then we choose $\phi_{i,\,i+1}=0$. It follows that $\Wt|_{\th\th}$ contains only $\psi_{i,i+1}$. Analysing the quiver one can see that there is only one term soaking up all fermionic zero modes and contributing to $Z$ thus the latter is non-zero. 
Similar argument holds for $k$ odd with a small modification. For $k$ odd we choose $\phi_{2,3}\neq 0$ also. This shows that $Z$ is non-zero at measure zero subset of the moduli space of the bosonic zero modes. By analyticity it is non-zero in some neighbourhood. Thus $Z$ is non-zero.

The remaining zero modes $\la$  we integrate as previously. The resulting contribution to the superpotential for the matter is
\beq
\d W~\propto~ h^k\Phi_{k,k+1}\Phi_{k+1,k}
\eeq

\section{Conclusions}

In this paper we have shown that multi-instanton configurations renormalize superpotentials. The mechanism responsible for this is lifting zero modes due to extra interaction occurring for intersecting instantons \cite{lifting}. We presented several explicit examples showing that the multi-instantons are generic phenomena.
One is tempted to formulate a hypothesis saying that superpotential is renormalized by all terms allowed by symmetries. The special treatment is required for anomalous 
$U(1)$ as it has been discussed in literature 
\cite{I-Uranga}.

Multi-instanton  contributions are suppressed by extra powers of $e^{-V}$, where $V$ corresponds to
 the volume of the cycle upon which the extra instantons wrap
 or by powers of the gauge coupling constant $h$ for string models realizing quiver gauge theories. 
 This may have phenomenological applications providing 
 new mechanism for building hierarchies between different terms of superpotentials. The mechanism is similar to the one used in \cite{MP}
 to produce Yukawa coupling hierarchies in heterotic string models.
 It may also lower the scale of the SUSY breaking in 
 models similar to \cite{aks} which we discussed in section \ref{conifold}. Although no phenomenologically viable model has been presented in this paper we hope that the discussed ideas are worth to  pursue.\\

{\bf Acknowledgments.}
The authors would like to thank Zygmunt Lalak and  Emilian Dudas,
Pablo Camara
 and other members of the theory group in Ecole Polytechnique for discussions and hospitality. \\
 
\appendix
\label{app-IIA}

\section{$T^6/(\ZZ_2\x\ZZ'_2)$ geometry}

The detailed description of geometric background we use in this paper can be found in \cite{frozen}, \cite{seesaw}, \cite{majorana}. Only necessary facts are recalled here. 
 
This compactification model contains one stack of four D6-branes wrapped on 3-cycle $\Pi_a$ given by $\Pi_a=(1,-1)\cdot(1,1)\cdot(1,1)$ (\cite{yukawa}). Tadpole cancellation condition as usually requires occurring of O6-planes which, in this geometry,  are homological to 
$\Pi_O=2(1,0)\cdot (0,1)\cdot (0,1)+2(1,0)\cdot (1,0)\cdot (1,0)-2(0,1)\cdot (0,1)\cdot (1,0)-2(0,1)\cdot (1,0)\cdot(0,1)$.  There also appears the image of D6-branes $\cR(\Pi_a)=\Pi'_a$ and due to orientifolding the gauge group is $U(4)_a=SU(4)_a\x U(1)_a$.  Moreover, we choose complex structure moduli of $T^6/(\ZZ_2\x\ZZ'_2)$ such that D$6_a$-branes are supersymmetric with respect to O6-planes.\\  
One can show that 
matter content of this model is formed by 32 chiral multiplets (\cite{majorana}) in antisymmetric $\bar{\bf 6}$ representation of $SU(4)_a$, which are localized at the intersections of D$6_a$- and D$6'_a$-branes:
\beq\label{chiDsix}
\Phi_{aa'}^I=(\phi_{aa'}^I,\psi_{aa'}^I,F^I_{aa'}),\quad I=1,\ldots,32;
\eeq
where we have suppressed Chan-Paton indices $i$, $j$.

For later convenience let us expose instantons present in described geometry:
\paragraph{$O(1)$ D2-instanton ($\widetilde{E}_1$)} This instanton is represented by 3-cycle $\widetilde{\Pi}_1=(1,0)\cdot(0,1)\cdot(0,-1)+\widetilde{\Pi}_{tw}$ (\cite{frozen}). We consider only configurations in which $\widetilde{E}_1$ is parallel to O6-plane but separated from it along the first two-torus $(T^2)^1$ and we take $\widetilde{\Pi}_{tw}$ as  `twisted' \cite{frozen} and invariant under orientifold action i.e.  $\cR(\widetilde{\Pi}_{tw})=\widetilde{\Pi}_{tw}$. Furthermore, $\widetilde{\Pi}_1$ has to  go through four fixed points of orbifold action in order to ensure rigidity.\\
Geometry of branes fixes the number of charged zero modes of $\widetilde{E}_1$, so one also need to know intersection numbers for $\widetilde{E}_1$ and D6-branes. These are given by $I_{1a}=1$ and $I_{1a'}=-1$.

\paragraph{$U(1)$ D2-instanton ($E_2$)} It corresponds to rigid 3-cycle $\Pi_2=(n^1,m^1)\cdot(n^2,m^2)\cdot(n^3,m^3)+\Pi_{tw}$, where $\Pi_{tw}$ is `twisted' as above. Likewise, the orientifold image $E'_2$ occupies $\Pi'_2=(n^1,-m^1)\cdot(n^2,-m^2)\cdot(n^3,-m^3)+\Pi'_{tw}$. As for D6-branes we fine-tune complex structure moduli such that $E_2$ is supersymmetric with respect to O$6$-planes\footnote{Hence with respect to D6-branes and $O(1)$ D2-instanton.}. Besides, we restrict our analysis to case $\Pi_{tw}\circ\widetilde{\Pi}_{tw}=0$ to simplify calculations. 
\subsection{Two extra instantons not intersecting D6-branes}\label{Atwonotint}
Let us impose the following conditions which realize geometry described in section \ref{twonotint}:
\bit
\item $E_2$ intersects $\widetilde{E}_1$ i.e. $I_{21}=\Pi_2\circ\widetilde{\Pi}_1\neq0$,
\item $E_2$ does not intersect D6-branes i.e. $I_{2a}=\Pi_2\circ\Pi_a=0=I_{2a'}=\Pi_2\circ\Pi_{a'}$,
\item $E_2$ intersects neither $E'_2$ nor O6-planes i.e. $I_{22'}=\Pi_2\circ\Pi'_{2}=0=I_{2O}=\Pi_2\circ\Pi_O$.
\eit
Using $(n^i,m^i)$ they can be written as:
\beqas
&&\mathrm{a)}\, m^1n^2n^3\neq0,\quad \mathrm{b)}\, n^1m^2m^3=0,\quad \mathrm{c)} (n^1+m^1)(n^2-m^2)(n^3-m^3)=0,\\
&&\mathrm{d)} (n^1-m^1)(n^2+m^2)(n^3+m^3)=0,\quad \mathrm{e)} -m^1n^2n^3-m^1m^2m^3+n^1n^2m^3+n^1m^2n^3=0.
\eeqas
It is straightforward to check that above equations are fulfilled by three families of instantons: 1) $(0,m^1)\cdot(\pm m^2,m^2)\cdot(n^3,\mp n^3)$, 2) $(\pm m^1,m^1)\cdot(n^2,0)\cdot(\pm m^3,m^3)$ and 3) $(\pm m^1,m^1)\cdot(\pm m^2,m^2)\cdot(n^3,0)$.  For each one it is easy to calculate the number of zero modes of $E_2$. Because cases 1), 2) and 3) are similar we shall consider only the first one. Let us define $l=I_{21}=m^1n^2n^3$ and distinct two situations $l>0$ and $l<0$ because the charges of zero modes $\Xi$ depend on the sign of $l$.
\medskip\\
\textbf{I)} $l>0$ - here one gets $l$ chiral multiplets $\Xi^M$ $(M=1,\ldots,l)$\\
The effective action for $E_2-\widetilde{E}_1$ system is just sum of \refeq{ee} and non-local interactions $g_e^I\la_i\Phi^I_{aa',ij}\la_j$, where $g_e^I$ is coupling constant for $I$-th field $\Phi^I_{aa'}$.

If $l=1$ then zero modes are\footnote{$x:=x_1=x_2$} $x$, $\th_1$, $\cM=\{\th_2\,,\thb_2\,,\psi\,,\psib\,,\phi\,,\phib\,,\la_i\}$ and path integral is given by\footnote{We set $q=1$ for simplicity.}:
\beqa\label{path}
\int d^4x\,d^2\th_1\,d\cM\,e^{-\widetilde{V}-V}e^{-g_e^I\la_i\Phi^I_{aa',ij}\la_j}e^{-(\phib\phi)^2}e^{\frac{i}{\sqrt{2}}(\phi\,\psib\,\thb_2-\phib\psi\,\th_2)},
\eeqa
where we have used equations of motion for auxiliary fields $D$ and $F^M$. $\widetilde{V}$ denotes volume of $\widetilde{\Pi}_1$ and $V$ is volume of $\Pi_2$. $\psi$, $\psib$ and $\th_2$ are soaked up by $(\phi\,\psib\,\thb_2)^2(\phib\psi\,\th_2)^2$ and integration over $\phi$ and $\phib$ produce non-zero constant thus \refeq{path} boils down to:
\beqa
&&\int d^4x\,d^2\th_1\,\a\,\mathrm{Pf}(g_e^I\Phi_{aa'}^I)e^{-\widetilde{V}-V},
\eeqa
where $\mathrm{Pf}(g_e^I\Phi_{aa'}^I)$ is Pfaffian of the matrix $g_e^I\Phi_{aa'}^I$ and $\a$ is non-zero constant. So all in all one gets the following contribution to the superpotential:
\beq\label{W}
\d W(\Phi,V,\widetilde{V})=\a\,\mathrm{Pf}(g_e^I\Phi^I_{aa'})e^{-\widetilde{V}-V}.
\eeq
\indent On the other hand if $l>1$ then there is too many modes $\Xi^M$ and one is not able to  soak up all of them just using $\phi^M\,\psib^{M'}\,\thb_2-\phib^M\psi^{M'}\,\th_2$. 

We can see that $\d W$ is generated only by such configurations which satisfy equation $m^1n^2n^3=1$. In case 1) it can be written as: $m^1m^2n^3=-1$. Thus altogether one gets 12 relevant two-instanton configurations.
\medskip\\ 
\textbf{II)} $l<0$ - $|l|$ chiral multiplets $\Xi^N$ ($N=1,\ldots,|l|$) appear in this situation\\
Here analysis is similar to that in \textbf{I)}. When $|l|=1$, then the contribution to the superpotential is given by \refeq{W} and if $|l|>1$ then path integral vanishes due to excess of fermionic zero modes and so does $\d W$. As above the condition $|l|=1=-m^1n^2n^3$ results in 12 non-trivial setups.

%
%


\end{document}